 \documentclass[aip,pop,reprint]{revtex4-1}
 \usepackage{dcolumn}
 \usepackage{bm}
 \usepackage{amsmath}
 \usepackage{tabularx}
\newcommand{\B}{\textrm{B}}

\newcommand{\D}{\textrm{D}}
\newcommand{\vc}{\mathbf}
\newcommand{\la}{\ensuremath{\langle}}
\newcommand{\ra}{\ensuremath{\rangle}}

\usepackage{graphicx}
\usepackage{color}

\begin{document}

% \preprint{UW-CPTC 09-5}

\title{Extending plasma transport theory to strong coupling through the concept of an effective interaction potential}


\author{Scott D.\ Baalrud}

%\thanks{}
\affiliation{Department of Physics and Astronomy, University of Iowa, Iowa City, IA 52242}
%\affiliation{Theoretical Division, Los Alamos National Laboratory, Los Alamos, New Mexico 87545}

\author{J\'{e}r\^{o}me Daligault}
\affiliation{Theoretical Division, Los Alamos National Laboratory, Los Alamos, New Mexico 87545}


\date{\today}

\begin{abstract}

A method for extending traditional plasma transport theories into the strong coupling regime is presented. Like traditional theories, this is based on a binary scattering approximation, but where physics associated with many body correlations is included through the use of an effective interaction potential. The latter is simply related to the pair-distribution function. Modeling many body effects in this manner can extend traditional plasma theory to orders of magnitude stronger coupling. Theoretical predictions are tested against molecular dynamics simulations for electron-ion temperature relaxation as well as diffusion in one component systems. Emphasis is placed on the connection with traditional plasma theory, where it is stressed that the effective potential concept has precedence through the manner in which screening is imposed. The extension to strong coupling requires accounting for correlations in addition to screening. Limitations of this approach in the presence of strong caging is also discussed. 

\end{abstract}

% 52.25.Dg   Plasma kinetic equations
% 52.25.Fi   Transport properties
% 52.27.Gr  Strongly-coupled plasmas
% 52.27.Lw  Dusty or complex plasmas
% 52.25.Kn Thermodynamics of plasmas

%\keywords{}

\maketitle


%%%%%%%%%
\section{Introduction\label{sec:intro}}

A central problem in the theory of strongly coupled plasmas is to describe how the many body physics of correlations affects transport properties. Traditional plasmas are so hot and dilute that the average particle kinetic energy greatly exceeds the potential energy of interaction. In this weakly coupled regime, interactions are long range and Coulomb collisions are accurately modeled as a series of small-angle binary scattering events.\cite{land:36} In contrast, the interaction potential energy exceeds the average particle kinetic energy in strongly coupled plasmas, so theory must account for both large angle collisions and many-body correlations. We recently proposed a method for extending the binary collision picture to strong coupling by accounting for correlations through an effective interaction potential.\cite{baal:13} In the present paper, this approach is developed further by detailing aspects of the analysis that were not included in the previous publication and by drawing connections with weakly coupled plasma theory.

This physically intuitive approach provides a practical means of extending the most commonly applied transport theories to stronger coupling. It may find application in ultracold plasmas,\cite{kill:07} dense plasmas,\cite{atze:04,chab:06} and dusty plasmas\cite{merl:04} where strong coupling effects can influence transport. It is especially advantageous for modeling systems in which the coupling strength can vary from weak to strong because it fits naturally within the framework of the Chapman-Enkog\cite{chap:70,ferz:72} or Grad\cite{grad,zhda:02} methods, providing the transport coefficients that the macroscopic fluid descriptions require as input.  The plasmas of interest are often multicomponent systems exhibiting a wide range of parameters. Fluid simulations typically demand that calls to routines providing transport coefficients be fast, such as a table lookup or evaluation of a closed-form equation. One approach is to build tables or best fit curves based on many ab-initio simulations, but this quickly becomes impractical for multicomponent plasmas when a large parameter space is required. Alternatively, approximate theories can be sought that balance accuracy and computational expense, while also providing insight into the physical processes at play.  Along these lines, we explore the extent to which the conventional binary scattering picture can be extended by using an effective interaction potential that includes correlation physics. 

The idea of imposing an effective interaction potential is not new. In fact, the approximation is used in traditional plasma theory as well, albeit in an indirect manner. Landau's seminal collision operator\cite{land:36} is based on a binary scattering picture where particles interact via their bare Coulomb interaction. However, this neglects the effect of screening and leads to an unphysical divergence. The effective potential concept is tacitly applied when screening is imposed through setting the maximum impact parameter to be the Debye length.  The screened Coulomb (Debye-H\"{u}ckel) potential has also been applied, but this does not include correlation effects that are important at strong coupling.\cite{libo:59,maso:67,paqu:86,khra:02,baal:12} In addition to our previous work on fluid transport,\cite{baal:13} recent computations of single particle stopping power have successfully included correlation effects using an effective potential concept.\cite{grab:13} 
%This concept has precedence in other theories as well, such as liquids (e.g., Lennard-Jones potentials)\cite{hans:06} and neutral gasses (e.g., hard sphere potentials),\cite{ferz:72} but it has not been widely applied to strongly coupled plasmas. 

The bulk of previous theoretical approaches either focus on deriving collision operators based on new closures of the BBGKY hierarchy,\cite{dali:09,ichi:92,dali:11} or calculate transport properties from higher-order equilibrium correlation functions.\cite{boer:82,hans:06} A popular approach to deriving new collision operators is to generalize the dielectric response function to include correlation physics through a local field correction. In this work, we take a different approach. We apply the standard Boltzmann collision operator with the only modification being the interaction force. To include correlation effects in this force, we draw on much previous work in the area of higher-order equilibrium correlation functions, including the concept of the potential of mean force and its connection with the pair distribution function [$g(r)$], as well as the hypernetted-chain (HNC) approximation.\cite{hans:06,hill:60} By equating the effective interaction potential with the potential of mean force, a closed set of equations is obtained from the HNC calculation of $g(r)$. Although we focus on classical HNC in this work, any other approximation for $g(r)$ could also be applied. For example, pair correlation functions in dense plasmas also require accounting for the quantum nature of electrons.\cite{saum:12} Utilizing the equilibrium relationship between $g(r)$ and the potential of mean force, the input to the theory becomes $g(r)$. 

In practice, evaluating transport coefficients using the effective potential theory consists of three steps. First the interaction potential must be specified. In this work, we obtain a model for the effective potential using the HNC approximation (Sec. \ref{sec:epot}), but any other means could also be applied.
Second, this potential is used to calculate the input functions for the transport theory.
For the Chapman-Enskog or Grad equations in particular, this requires solving three imbedded integrals [Eqs. (\ref{eq:theta}), (\ref{eq:xilk}) and (\ref{eq:sigl})] for the scattering angle, momentum-transfer cross-section and $\Omega$-integrals.
Third, the $\Omega$-integrals provide the input for calculating various transport coefficients in these theories. These are typically evaluated as functions of the coupling strength $\Gamma = e^2Z^2/(ak_BT)$ where $a=(3/4\pi n)^{1/3}$ is the inter-particle spacing. We provide the theoretical description in general terms valid for multicomponent systems, but apply it only to the specific system of a classical one-component plasma (OCP) in this paper. 

This paper is organized as follows. The basic transport models are discussed in Sec.~\ref{sec:trans}. Application to Chapman-Enskog and Grad theories is emphasized where the various transport coefficients are determined from a matrix of integral relations called the $\Omega$-integrals. Section~\ref{sec:epot} discusses how the interaction potential is obtained from equilibrium properties of the BBGKY hierarchy and the HNC approximation.  Section~\ref{sec:omega} discusses numerical evaluation of the $\Omega$-integrals. Section~\ref{sec:compare} shows a comparison between the calculation and molecular dynamics (MD) simulations for diffusion in one-component plasma (OCP), and the electron-ion temperature relaxation rate. Limitations of the theory and possible extensions are discussed in Sec.~\ref{sec:lim}. 

%%%%%%%%
\section{Transport models \label{sec:trans}}

We start from the binary collision approximation, which is the basis of the Boltzmann collision operator\cite{ferz:72}
\begin{equation}
C_B^{s-s^\prime} = \int d^3v^\prime d\Omega\, \sigma_{ss^\prime} u [ f_s(\hat{\vc{v}}) f_{s^\prime} (\hat{\vc{v}}^\prime) - f_s (\vc{v}) f_{s^\prime} (\vc{v}^\prime)] \label{eq:boltz}
\end{equation}
where $(\vc{v}, \vc{v}^\prime)$ are the initial velocities of colliding particles and $(\hat{\vc{v}}, \hat{\vc{v}}^\prime)$ are the post-collision velocities ($s$ and $s^\prime$ denote species). Here, $\vc{u} = \vc{v} - \vc{v}^\prime$ is the relative velocity vector ($u = |\vc{u}|$), $\sigma_{ss^\prime}$ is the differential scattering cross section and $d \Omega = d\phi d\theta \sin \theta$ is the solid angle. Equation~(\ref{eq:boltz}) provides the basis for kinetic and transport theories in a variety of disciplines. The input is the differential scattering cross section, which is commonly determined by classical dynamics once the interaction force is specified. One difficulty in applying Eq.~(\ref{eq:boltz}) is the dependence on the distribution functions evaluated at the post-collision velocities ($\hat{\vc{v}}, \hat{\vc{v}}^\prime$). In traditional plasma physics, this is circumvented by applying the small scattering angle expansion: $\hat{\vc{v}} = \vc{v} + \Delta \vc{v}$ where $\Delta \vc{v} \ll \vc{v}$. This expansion leads to the familiar plasma collision operators of the Landau\cite{land:36} or Fokker-Planck\cite{rose:57} forms. Transport theories, such as Braginskii's,\cite{brag:65} are formulated by expanding the resultant weakly coupled collision operator about equilibrium. 

The small scattering angle expansion is not viable for strongly coupled plasmas because scattering angles are large in these systems. Although the full Boltzmann collision operator must be dealt with at the kinetic level, transport coefficients for fluid equations are determined from velocity moments of the form $\la \chi \ra^{s-s^\prime} = \int d^3v \chi_s (\vc{v}) C_B^{s-s^\prime}$ where $\chi_s$ are velocity-dependent functions related to continuity, momentum, energy, etc.\ ($m_s, m_s\vc{v}, m_sv^2, \ldots$). For the Boltzmann collision operator, symmetry properties of the binary scattering process can be exploited to write these integrals in a form where the post-collision velocity vectors come outside of the distribution functions\cite{baal:12}
\begin{equation}
\la \chi \ra^{s-s^\prime} = \int d^3v d^3v^\prime \lbrace \Delta \chi_s \rbrace f_s (\vc{v}) f_{s^\prime}(\vc{v}^\prime) . \label{eq:chi}
\end{equation}
Here, $\lbrace \Delta \chi_s \rbrace = \int d\Omega\, \sigma_{ss^\prime} u \Delta \chi_s$ and $\Delta \chi_s = \chi_s (\hat{\vc{v}}) - \chi_s (\vc{v})$. 

Transport coefficients of the form in Eq.~(\ref{eq:chi}) are simplified by applying conservation laws to write $\Delta \chi_s$ in terms of $\vc{v}$ and $\vc{v}^\prime$. For instance, the momentum equation in multi-fluid descriptions depends on the friction force density $\vc{R}^{s-s^\prime} = \la m_s \vc{v} \ra^{s-s^\prime}$. Applying conservation of momentum to Eq.~(\ref{eq:chi}) leads to\cite{baal:12} 
\begin{equation}
\vc{R}^{s-s^\prime} = m_{ss^\prime} \int d^3u\, \lbrace \Delta \vc{u} \rbrace \int d^3v^\prime\, f_{s} (\vc{u} + \vc{v}^\prime) f_{s^\prime} (\vc{v}^\prime),  \label{eq:r2}
\end{equation}
where $\Delta \vc{u} = u [\sin \theta \cos \phi \hat{x} + \sin \theta \sin \phi \hat{y} - 2 \sin^2(\theta/2) \hat{u}]$. Likewise, the energy conservation equation relies on the energy exchange density $Q^{s-s^\prime} = \la \frac{1}{2} m_s v^2 \ra - \vc{V}_s \cdot \vc{R}^{s-s^\prime}$. Applying conservation of energy to Eq.~(\ref{eq:chi}) provides\cite{baal:12}
\begin{equation}
Q^{s-s^\prime} = m_{ss^\prime} \int d^3u\, \lbrace \Delta \vc{u} \rbrace \cdot \vc{I}_\vc{u} , \label{eq:q2}
\end{equation}
where $\vc{I}_{\vc{u}} = \int d^3v^\prime ( \vc{v}^\prime - \vc{V} + m_{ss^\prime} \vc{u}/m_{s^\prime}) f_{s} (\vc{u} + \vc{v}^\prime)  f_{s^\prime} (\vc{v}^\prime)$ and $m_{ss^\prime} =m_s m_{s^\prime}/(m_s + m_{s^\prime})$ is the reduced mass.

Evaluating these transport coefficients requires a scattering cross section and a model for the distribution functions. For the scattering cross section, it is convenient to write the $\lbrace \Delta \chi_s \rbrace$ integrals in terms of the impact parameter $b$ and the scattering angle $\theta = \pi - 2\Theta$ by applying the substitution $d\Omega \sigma_{ss^\prime} = bdbd\phi$. The scattering angle is determined from the elementary classical dynamics of two particles $s$ and $s'$ interacting through a central force $-\nabla \phi_{ss^\prime}$, which provides 
\begin{equation}
\Theta = b \int_{r_o}^\infty dr\, r^{-2} \biggl[ 1 - \frac{b^2}{r^2} - \frac{2\phi_{ss^\prime} (r)}{m_{ss^\prime} u^2} \biggr]^{-1/2} . \label{eq:theta}
\end{equation}
Here, $r_o$ is the distance of closest approach, which is determined from the largest root of the denominator in Eq.~(\ref{eq:theta}). 

The general formulation of Eqs.~(\ref{eq:chi})-(\ref{eq:theta}) applies to any model for the distribution functions, but the most commonly applied model is a perturbation about equilibrium. For instance, the transport coefficients in Grad's moment method can be derived by applying this expansion to Eqs.~(\ref{eq:r2}), (\ref{eq:q2}) and similar higher-order moments.\cite{grad,zhda:02} The resulting transport coefficients can be written in terms of integrals of the form
\begin{equation}
\Omega_{ss^\prime}^{(l,k)} = \sqrt{\pi} \bar{v}_{ss^\prime} \int_0^\infty d\xi \xi^{2k+3} e^{-\xi^2} \int_0^\pi  d\theta \sigma_{ss^\prime} \sin \theta  (1 - \cos^l \theta) 
\end{equation}
in which $\xi = u/\bar{v}_{ss^\prime}$, $\bar{v}_{ss^\prime}^2 = v_{Ts}^2 + v_{Ts^\prime}^2$ and $v_{Ts}^2 = 2T_s/m_s$. For example, the 13N moment form of Grad's equations are provided in terms of $\Omega$-integrals in Zhdanov.\cite{zhda:02} Alternatively, the Chapman-Enskog approach also specifies various transport coefficients, such as diffusivity, conductivity, viscosity, etc., in terms of $\Omega$-integrals.\cite{chap:70,ferz:72} We focus on calculating the $\Omega$ integrals from the effective potential as these provide the input to these theories. 

To facilitate the connection with weakly coupled plasma theory, the $\Omega$-integrals can alternatively be written in the form 
\begin{equation}
\Omega_{ss^\prime}^{(l,k)} = \frac{3}{16} \frac{m_s}{m_{ss^\prime}} \frac{\nu_{ss^\prime}}{n_{s^\prime}} \frac{\Xi_{ss^\prime}^{(l,k)}}{\Xi_{ss^\prime}} , \label{eq:cints}
\end{equation}
where 
\begin{equation}
\Xi_{ss^\prime}^{(l,k)} = \frac{1}{2} \int_0^\infty d\xi\, \xi^{2k+3} e^{-\xi^2} \bar{\sigma}_{ss^\prime}^{(l)} / \sigma_o \label{eq:xilk}
\end{equation}
is a ``generalized Coulomb logarithm'' associated with the $(l,k)^\textrm{th}$ collision integral. We apply this name in this context because $\Xi^{(1,1)} \rightarrow \ln \Lambda$ and $\Xi^{(l,k})/\Xi^{(1,1)} \rightarrow$ constants (independent of $\Lambda$) in the weakly coupled limit. Here
\begin{equation}
\nu_{ss^\prime} \equiv \frac{16 \sqrt{\pi} q_s^2 q_{s^\prime}^2 n_{s^\prime}}{3 m_s m_{ss^\prime} \bar{v}_{ss^\prime}^3} \Xi_{ss^\prime} \label{eq:nu}
\end{equation}
is a reference collision frequency,
\begin{equation}
\bar{\sigma}_{ss^\prime}^{(l)} = 2 \pi \int_0^\infty db\, b [1 - \cos^{l} (\pi - 2 \Theta) ]  \label{eq:sigl}
\end{equation}
is the $l^{\textrm{th}}$ momentum-transfer cross section and $\sigma_o = (\pi q_s^2 q_{s^\prime}^2)/(m_{ss^\prime}^2 \bar{v}_{ss^\prime}^4)$ is a reference cross section.

The only input to these equations is the interaction potential energy $\phi_{ss^\prime}(r)$. Our approach uses an effective potential that self-consistently includes screening and correlation physics. The next section discusses a theoretical basis for this potential and a means of calculating it.

%%%%%%%%%
\section{Effective interaction potential \label{sec:epot}}

For convenience, in this section, we consider a one-component system and drop the subscipts $s,s'$ denoting the species. However, the arguments can be generalized to multicomponent systems. 

%%%%%%%%%
\subsection{Potential of mean force}

In this section, we establish a connection between the effective interaction potential energy $\phi(r)$ (hereinafter referred to as the ``effective potential'' for brevity) and the pair distribution function $g(r)$. The pair distribution function is related to the probability, $\rho^{(2)}(\vc{r}, \vc{r}') d\vc{r}d\vc{r}'$, of finding two particles of the system in an elementary volume element $d\vc{r}d\vc{r}'$ around $(\vc{r},\vc{r}')$, irrespective of the positions of the other particles and irrespective of all velocities, with
\begin{eqnarray}
\rho^{(2)}(\vc{r}, \vc{r}')&=&\left\langle \sum_{i=1}^{N}{\sum_{j=1,j\neq i}^{N}{\delta({\bf r}-{\bf r}_i) \delta({\bf r}'-{\bf r}_j)}}\right\rangle_{eq}\nonumber\\
&=& \frac{N(N-1)}{{\cal{Z}}} \int e^{-U/k_B T} d\vc{r}_3\ldots d\vc{r}_N  . \label{eq:g2int}
\end{eqnarray}
Here, ${\cal Z} \equiv \int \exp(-U/k_BT) d\vc{r}_N$ is the configurational integral and $U \equiv \sum_{i,j}{v(|\vc{r}_i - \vc{r}_j|)}$, where $v$ is the bare interaction potential energy.\cite{hill:60} For a homogeneous and isotropic system, such as considered in this paper, $\rho^{(2)}(\vc{r}, \vc{r}')$ is a function only of the separation $|{\bf r}-{\bf r}'|$. The pair distribution function is defined as
\begin{equation}
n^2g(r) =\rho^{(2)}(r) \label{eq:rhogr}
\end{equation}
where the thermodynamic limit is implicitly understood. Physically, $ng(r)$ represents an average radial density distribution around individual particles. For illustration, Fig.~\ref{fg:gr_r} shows $g(r)$ for a OCP at different coupling strengths.
At strong coupling, density oscillations are produced by nearest neighbor particles: the number of particles lying within the distance $r$ to $r+dr$ from a given particle is $4\pi r^2 n g(r)dr$ and the peaks in $g(r)$ represent shells of neighbors around that particle. At weak coupling, the local density around a reference particle is everywhere equal to its average $n$, i.e. $g(r)=1$, except in its immediate neighborhood due to the strong inter-particle Coulomb repulsion.

The pair distribution function can be related to an interaction potential using Eq.~(\ref{eq:g2int}). In particular, if we define $\phi$ by
\begin{equation}
g(r) = \exp [-\phi(r)/k_\B T] , \label{eq:grphi}
\end{equation}
Eq.~(\ref{eq:g2int}) can be written in the form\cite{hill:60}
\begin{equation}
- \nabla \phi =   \frac{\int e^{-U/k_\B T} (-\nabla_1 U)d \vc{r}_3 \ldots d \vc{r}_N}{\int e^{-U/k_\B T} d \vc{r}_3 \ldots d \vc{r}_N} .  \label{eq:epdef}
\end{equation}
Thus, we see that the force ($-\nabla \phi$) in this relationship is that acting on particle 1, taking particles 1 and 2 held at fixed positions ($\vc{r}_1$ and $\vc{r}_2$) and averaging over the positions of all other particles.
This is precisely what is desired for considering the binary collision between particles 1 and 2, accounting for average effects of the medium. Thus, we equate this so-called ``potential of mean force'' $\phi$ with the effective interaction potential and use Eq.~(\ref{eq:grphi}) to relate this to the pair distribution function. A variety of techniques have been developed to approximate $g(r)$ under various conditions. In the next section, we will consider the HNC approximation. 
%Thus, $g(r)$ becomes the input to the transport theory. 

However, we first remark that while plasma kinetic theories do not typically make explicit reference to a pair distribution function, one is nevertheless implicit in each theory.
In order to draw a connection with these traditional theories, consider the second equation of BBGKY for a classical one-component plasma
\begin{subequations}
\begin{align}
&\frac{\partial g_2(1,2)}{\partial t} = \left[L_1^0+L_2^0\right]g_2(1,2)+L_{12}f(1)f(2)   \label{1}\\ 
&+L_{12}g_2(1,2)   \label{l2}\\
&+\int{d3\left[ L_{13}f(1)g_2(2,3)+L_{13}f(3)g_2(1,2) + (1\leftrightarrow 2) \right]}   \label{l3}\\ 
&+\int{d3 (L_{13}+L_{23})g_3(1,2,3)}   \label{l4} 
\end{align}
\label{BBGKY2}
\end{subequations}
in which $L_i^0 = -\frac{\vc{p}_i}{m} \cdot \nabla_i$, $L_{i,j} = \nabla v(|{\bf r}_i-{\bf r}_j|)\cdot \left(\partial /\partial {\bf p}_i - \partial / \partial {\bf p}_j\right)$, $f$ is the distribution function at time $t$, $g_n$ is the $n$-particle correlation function at time $t$, and $1=({\bf r}_1,{\bf p}_1)$. The usual plasma kinetic theories (namely the Landau,\cite{land:36} the Boltzmann\cite{ferz:72} and the Lenard-Balescu\cite{lena:60} equations) can be obtained by neglecting certain terms in Eq.~(\ref{BBGKY2}) in order to close the infinite hierarchy of equations between the $g_n$'s.\cite{libo:98,dali:11} These closures rely on a systematic ordering of the correlations $g_n=O(\alpha^m)$ (m integer) in terms of an adequately chosen small parameter $\alpha$. The resulting equation for $g_2$ is then solved analytically in terms of $f$ and the solution is introduced in the first BBGKY equation to derive a kinetic equation for $f$ that is of second order in $\alpha$; additional hypotheses are made\cite{libo:98} to produce a tractable equation, in particular the ``Markovianization'' of the collision operator. For our purposes, it is instructive to consider the equilibrium limit of the approximate equation for $g_2$ obtained by setting $\partial_t g_2(1,2) =0$. The solution is related to the pair-distribution function
\begin{eqnarray}
g_2^{eq}(1,2)=n^2f_{M}({\bf p}_1)f_{M}({\bf p}_2)\big[g(|{\bf r}_1-{\bf r}_2|)-1\big] \label{g2eq}
\end{eqnarray}
where $f_M$ is a Maxwellian.

The Boltzmann equation describes dilute neutral gases with an arbitrarily strong interaction potential. In this case, the expansion parameter is $\alpha=n l_{C}^{3}$ where $n$ is the particle density and $l_{C}$ is the correlation length (assumed finite), and the Boltzmann closure is obtained by neglecting (\ref{l3}) and (\ref{l4}).
Here $l_C$ represents the maximum value of the range of the correlation function $g_n$; e.g., if $|{\bf r_1}-{\bf r_2}|> l_C$, then $g_n({\bf r}_1,{\bf r}_2,\dots,{\bf r}_n)=0$. The equilibrium limit of this closure provides Eq.~(\ref{g2eq}) with
\begin{eqnarray}
g(r)=e^{-v(r)/k_BT}\quad\text{Boltzmann}\,.
\end{eqnarray}
Not surprisingly, the $g(r)$ of the Boltzmann closure equals the Boltzmann factor as would be obtained by looking at the density around an impurity in an ideal gas: the potential of mean-force equals the bare potential, $\phi=v$. 

The Landau closure is valid for weakly coupled systems and $\alpha$ is the dimensionless strength of the potential; $v(r)=\alpha \bar v(r)$ where $\bar v(r)$ is dimensionless of $\mathcal{O}(1)$ for all $r$. It is obtained by neglecting terms (\ref{l2}), (\ref{l3}) and (\ref{l4}) in Eq.~(\ref{BBGKY2}). In the equilibrium limit, we find
\begin{eqnarray}
g(r)=1-v(r)/k_BT\quad\text{Landau}\,.
\end{eqnarray}
This corresponds to the Boltzmann factor in the high-temperature limit $v(r)/k_BT\ll 1$.

Finally, the Lenard-Balescu-Guernsey equation\cite{lena:60} is an equation for weakly-coupled plasmas in which the collisional interactions occur via an effective, dynamically screened potential. The parameter $\alpha$ is the particle charge $q^2$ together with the condition that $nq^2$ remains finite.\cite{libo:98,lena:60} This is satisfied whenever the coupling strength is small $q^2 n^{1/3}/k_{B}T\ll 1$; then the closure neglects (\ref{l2}) and (\ref{l4}). The term (\ref{l2}) kept in the Lenard-Balescu equation but dropped in the Boltzmann equation allows for the renormalization of the bare Coulomb interaction into a dynamically screened two-body interaction between the particles of the plasma.  In the equilibrium limit, this provides Eq.~(\ref{g2eq}) with
\begin{eqnarray}
g(r)=1-v_\textrm{sc}(r)/k_BT \quad\text{Lenard-Balescu} \label{grLB}
\end{eqnarray}
where $v_\textrm{sc}=q^2e^{-r/\lambda_D}/r$ is the screened Coulomb (Debye-H\"uckel) potential. Equation (\ref{grLB}) can be regarded as an approximation of Eq.~(\ref{eq:grphi}) with effective potential $\phi(r)=v_\textrm{sc}(r)$ in the limit $\phi(r)/k_BT\ll 1$. In contrast, the contribution (\ref{l3}) kept in the Boltzmann approximation but discarded in the Lenard-Balescu equation, is responsible for the bare two-particle interactions (large-angle scattering) describing close encounters in a dilute gas. A common way to extend the Lenard-Balescu approach is to modify the dielectric response function to account for close interaction physics using a local field correction obtained, in part, from $g(r)$.\cite{ichi:92,dali:09,bene:12}

The traditional plasma closures neglect the correlation functions $g_n$ with $n\geq 3$ [term (\ref{l4})], which is inappropriate when many-body correlations are moderate or strong. Their extension to the moderately and strongly coupled regimes confronts the full complexity of the many-body problem. Both from the practical and the physical standpoint, it would be quite useful to develop a kinetic theory that retains the desirable properties of the Boltzmann equation, whereby particles interact via an effective binary potential that integrates the average effects of the surrounding medium on the bare interactions. While progress has been made over the years to tackle this formidable problem (e.g., see Ref.~\onlinecite{dali:11}), the state-of-the-art solutions are still not tractable for practical calculations yet. In the present work, instead of deriving a kinetic equation from first principles and calculating the resulting transport properties, we phenomenologically include correlation and screening physics in the conventional binary scattering picture. The binary collision picture phenomenology leads to a Boltzmann collision operator without necessarily specifying the interaction force, outside of the requirement that it be central and conservative.\cite{ferz:72}  We use the HNC approximation to calculate accurate pair-distribution functions $g(r)$ that include both screening and correlation effects self-consistently. For a given $g(r)$, we obtain the effective potential $\phi(r)$ according to Eq.~(\ref{eq:grphi}) and use the results of Sec.~\ref{sec:trans} to calculate the transport properties. As we will see, this approach reduces to the popular approximations discussed above in the appropriate limits. In addition, it avoids the unphysical divergences encountered when one naively uses the bare Coulomb potential in the collision integrals, and it takes proper account of the correlated collisions characteristic of strongly coupled systems.

%%%%%%%%%
\subsection{Calculating the effective potential from HNC} 

The previous section established a relationship between the effective interaction potential and the pair distribution function [Eq.~(\ref{eq:grphi})]. Thus, any theory or experiment that provides $g(r)$ also provides the input for the effective potential theory. In practice, of course, it is desirable to use an approximation that does not rely on computationally expensive simulations. Here, we consider the HNC closure as an approximation for obtaining a $g(r)$ that includes correlation physics.

HNC is a well-established model for the one-component plasma. As described in detail in Ref.~\onlinecite{hans:06}, it can be derived from equilibrium statistical mechanics using advanced perturbative methods, or from density functional theory. In this approximation, $g(r)$ is determined from the coupled set of equations\cite{hans:06}
\begin{subequations}
\begin{eqnarray}
g(\vc{r})&=&\exp [- v(\vc{r})/k_\B T + h(\vc{r}) - c(\vc{r}) ]\label{eq:hncgr}\\
\hat{h}(\vc{k})&=&\hat{c}(\vc{k})[ 1 + n \hat{h}(\vc{k})] \,,  \label{eq:hnchk} 
\end{eqnarray}
%\label{hncequations}
\end{subequations}
where $h(r)=g(r)-1$ is the pair-correlation function and $\hat{h} (\vc{k})$ denotes the Fourier transform of $h(\vc{r})$. Equation (\ref{eq:hnchk}) is an exact relation known as the Ornstein-Zernike (OZ) relation. Equations (\ref{eq:hncgr}) and (\ref{eq:hnchk}) can be efficiently solved iteratively starting from a reasonable guess for $c({\bf r})$ [e.g., $c({\bf r})=-v({\bf r})/k_BT$].

\begin{figure}
\begin{center}
\includegraphics[width=8cm]{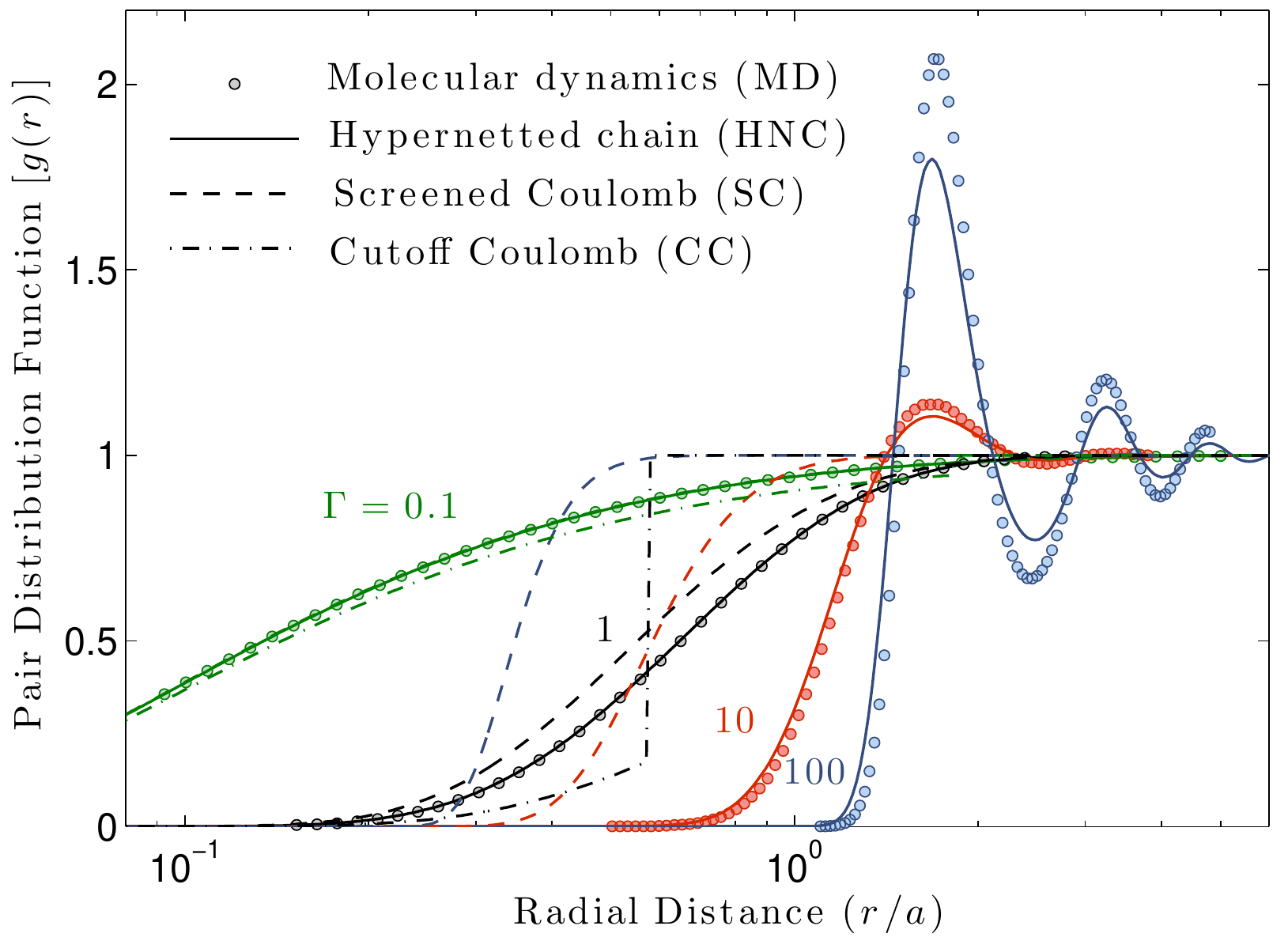}
\caption{Pair distribution function for a OCP obtained from MD (circles), HNC (solid lines), screened Coulomb potential (dashed lines) and the cutoff Coulomb potential (dash-dotted lines - shown only for $\Gamma = 0.1$ and 1).}
\label{fg:gr_r}
\end{center}
\end{figure}

\begin{figure}
\begin{center}
\includegraphics[width=8cm]{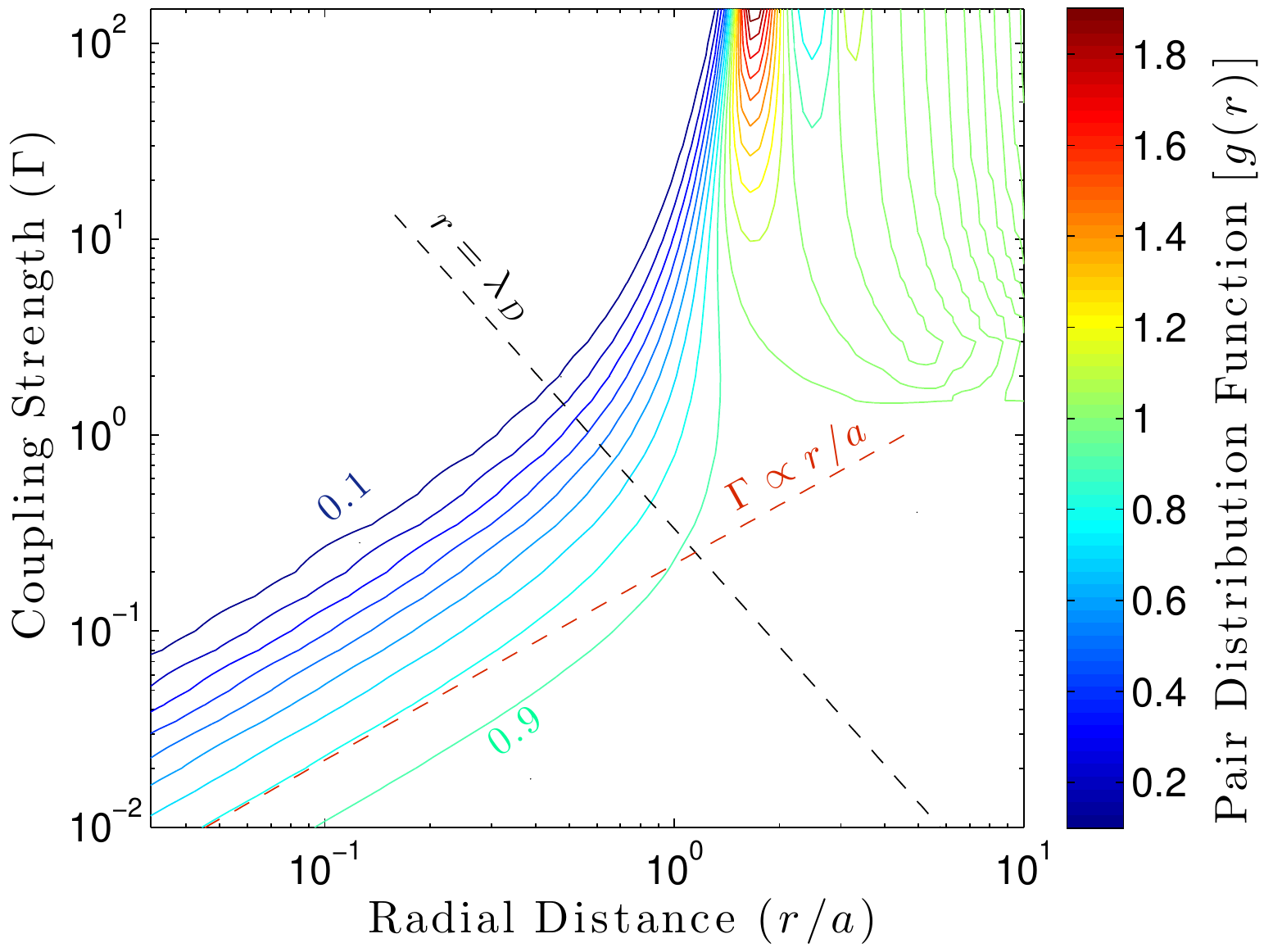}
\caption{Lines of constant $g(r)$ obtained from the HNC approximation for a OCP. Lines start at 0.1 and have a linear spacing in increments of 0.1. This shows the screened Coulomb behavior at weak coupling and correlation effects at strong coupling. The $\Gamma \propto r/a$ curve represents scaling of the bare Coulomb potential $v = \Gamma/(r/a)$.}
\label{fg:gr_cont}
\end{center}
\end{figure}

The HNC equation (\ref{eq:hncgr}) is an approximation of the exact relation $g(\vc{r})=\exp [- v(\vc{r})/k_\B T + h(\vc{r}) - c(\vc{r})+b({\bf r})]$ obtained by neglecting the so-called bridge function $b({\bf r})$. Very good analytical expressions for $b({\bf r})$ obtained by interpolation of exact numerical data can be found in the literature\cite{Iyet:86} to calculate $g(r)$'s across coupling regimes. In what follows, we work with the conventional HNC approximation (\ref{eq:hncgr}). 

Comparing Eq.~(\ref{eq:hncgr}) and Eq.~(\ref{eq:grphi}), we find that the effective potential is
\begin{eqnarray}
\phi(r)&=&v(r)-k_BT\left[h(r)-c(r)\right]\nonumber\\
&=&v(r)-nk_BT\int{d{\bf r}' c(|{\bf r}-{\bf r}'|)h({\bf r}')}\label{phiHNC}\,
\end{eqnarray}
in the HNC approximation. Equation (\ref{phiHNC}) expresses the effective potential between two particles at a distance $r$ apart as the sum of the bare interaction $v(r)$ plus a term that describes the effect of the surrounding medium of the latter. In the weakly coupled regime, as discussed below, one recovers the Debye screened potential.

The Debye screened potential is routinely derived in traditional plasma physics from Poisson's equation and the Boltzmann density relation, or from the linear dielectric response function.\cite{nich:83} It can also be obtained from the weakly coupled limit of the HNC approximation. Equating (\ref{eq:grphi}) and (\ref{eq:hncgr}) gives 
\begin{equation}
c(r) = \exp \biggl( - \frac{\phi(r)}{k_BT}\biggr) - 1 + \frac{\phi(r)}{k_BT} - \frac{v(r)}{k_BT}  
\end{equation}
which reduces to $c(r) \simeq - v(r)/k_BT$ in the weakly coupled limit ($|\phi| \ll k_BT$). This implies $h(r) = -\phi(r)/k_BT$. The OZ relation (\ref{eq:hnchk}) can then be written $\hat{\phi}(\vc{k}) = \hat{v}(\vc{k})/(1 + n \hat{v}(\vc{k})/k_BT)$. If the the bare interaction is taken to be Coulombic $\hat{v}(\vc{k}) = 4\pi q^2/k^2$, this reduces to the screened Coulomb potential
\begin{equation}
\phi(r)/k_BT = v_{\textrm{sc}}(r)= (q^2/r) \exp (-r/\lambda_\D)  , \label{eq:sc}
\end{equation} 
where $\lambda_\D=\sqrt{k_BT/4\pi q^2n}$ is the Debye length.

Figure~\ref{fg:gr_r} shows a comparison between the $g(r)$ obtained from MD simulations (described below), HNC, screened Coulomb potential and cutoff bare Coulomb potentials. The four descriptions all converge in the weakly coupled limit. Screened Coulomb remains accurate up until correlation effects onset at $\Gamma \gtrsim 1$. Beyond this point, it does not capture the correct interaction length scale, or the oscillatory behavior characteristic of correlation effects. The HNC approximation remains sufficiently accurate for our purposes over this entire range of coupling strength (up to $\Gamma$ of 100).

Figure~\ref{fg:gr_cont} shows a contour plot of $g(r)$ obtained from the HNC approximation at various coupling strengths. The lines in this figure show constant $g(r)$ contours starting at $0.1$ with a spacing of $0.1$.  The red dashed line corresponds to the scaling of the bare Coulomb interaction $v (r) =  \Gamma /(r/a)$. This shows that for weakly coupled plasmas, the interaction potential is well characterized by a bare Coulomb up until $r = \lambda_\D$, which is shown as a dashed line in the figure. After this point, $g(r)$ rapidly asymptotes to $1$, signifying an effective truncation of the interaction potential. This sharp transition provides a visualization of Debye screening. For large $\Gamma$, the behavior is fundamentally different. Here the interaction length scale is characterized by the inter-particle spacing, rather than the Debye length. It also shows the onset of oscillations associated with nearest neighbor potential wells at strong coupling.

%%%%%%%%%
\section{Evaluating the $\Omega$ integrals \label{sec:omega}}

In this section, we calculate the $\Omega$-integrals based on each of these models for $g(r)$ in order of increasing complexity. The discussion of the cutoff and screened Coulomb potentials are written in multicomponent form, but the HNC discussion concentrates on the OCP.

%%%%%%%%%
\subsection{Cutoff Coulomb potential}

In this section, a cutoff Coulomb potential is considered in an effort to make an explicit connection between the effective potential theory and Landau's seminal approach, which imposed screening by limiting the impact parameter to be within a Debye length. The formulation in Sec.~\ref{sec:trans} is based on an effective potential that is a function of the distance between colliding particles, with no reference to impact parameter. Here we show that an equivalent formulation can be made in the weakly coupled limit by truncating the interaction distance at the Debye length
\begin{eqnarray}
\phi_{ss^\prime} (r) =  \label{eq:phitrun}
\left\lbrace \begin{array}{ll}
q_s q_{s^\prime}/r , & \textrm{if}\ r<\lambda_\D \\
0,  & \textrm{if}\ r > \lambda_\D .
\end{array} \right. 
\end{eqnarray}

Applying Eq.~(\ref{eq:phitrun}) to calculate the scattering angle in Eq.~(\ref{eq:theta}) gives
\begin{eqnarray}
\Theta =  \label{eq:thetatrun}
\left\lbrace \begin{array}{ll}
\textrm{arccos} \biggl( \frac{b/\lambda_\D + \kappa/b}{\sqrt{1 + \kappa^2/b^2}} \biggr) + \textrm{arcsin} \biggl(\frac{b}{\lambda_\D}\biggr) , & r_o<\lambda_\D \\
\pi/2,  & r_o > \lambda_\D .
\end{array} \right. 
\end{eqnarray}
where 
\begin{equation}
r_o = \frac{b}{- \kappa/b + \sqrt{\kappa^2/b^2 + 1}}
\end{equation}
is the distance of closest approach and $\kappa = q_sq_{s^\prime}/(m_{ss^\prime}u^2)$. In the limit $b/\lambda_\D \ll \kappa/b$ and $\kappa/\lambda_D \ll 1$, this reduces to the Rutherford scattering angle with Landau's impact parameter cutoff 
\begin{eqnarray}
\Theta =  \label{eq:thetaland}
\left\lbrace \begin{array}{ll}
\textrm{arccos} \biggl( \frac{\kappa/b}{\sqrt{1 + \kappa^2/b^2}} \biggr)  , & b<\lambda_\D \\
\pi/2,  & b > \lambda_\D .
\end{array} \right. 
\end{eqnarray}

Applying Eq.~(\ref{eq:thetaland}) to the momentum-transfer cross section from Eq.~(\ref{eq:sigl}) gives
\begin{equation}
\frac{\bar{\sigma}_{ss^\prime}^{(l)}}{\lambda_{\D}^2} = 2\pi \int_0^1 d \bar{b}\, \bar{b} \biggl\lbrace 1 - \biggl[ \frac{\bar{b}^2 - 1/(\Lambda \xi^2)^2}{\bar{b}^2 + 1/(\Lambda \xi^2)^2} \biggr]^l \biggr\rbrace   \label{eq:siglcc}
\end{equation}
where $\bar{b} = b/\lambda_\D$ and $\Lambda = m_{ss^\prime} \bar{v}_{ss^\prime}^2 \lambda_\D/|q_s q_{s^\prime}|$ is the plasma parameter. After writing the term in braces using a common denominator and expanding the numerator for $\bar{b}^2 \gg 1/(\Lambda \xi^2)^2$ the momentum-transfer cross section reduces to
\begin{equation}
\frac{\bar{\sigma}_{ss^\prime}^{(l)}}{\lambda_\D^2} \simeq 4\pi l \frac{\ln (\Lambda \xi^2 )}{(\Lambda \xi^2)^2}  \label{eq:sigiwc}
\end{equation}
in the weakly coupled limit. Applying this to Eq.~(\ref{eq:xilk}) gives the following closed form expression for the generalized Coulomb logarithms 
\begin{equation}
\Xi_{ss^\prime}^{(l,k)} = l \Gamma (k) \ln \Lambda  
\end{equation}
in the weakly coupled limit. Here $\Gamma(k)$ is the Gamma function. 

Convergent forms of these expressions can be obtained by integrating Eq.~(\ref{eq:siglcc}) directly. For example, the first momentum-transfer cross section is 
\begin{equation}
\frac{\bar{\sigma}_{ss^\prime}^{(1)}}{\lambda_\D^2}  = \frac{2\pi}{(\Lambda \xi^2)^2} \ln (1 + \Lambda^2 \xi^4)  .
\end{equation}
Applying the approximation\cite{khra:02} $\ln (1 + \Lambda^2 \xi^4) \simeq 2 \ln (1 + \Lambda \xi^2)$, a simple convergent expression for the lowest order Coulomb logarithm can be obtained\cite{baal:12}
\begin{equation}
\Xi_{ss^\prime}^{(1,1)} \simeq \exp(\Lambda^{-1}) E_1 (\Lambda^{-1})
\end{equation}
where $E_1$ is the exponential integral. Although this expression is convergent, the weakly coupled expansion has been applied throughout this section, which restricts its applicability. However, the expansion procedure can be used to capture order unity corrections to the Coulomb logarithms. Expanding the analytic solutions of Eq.~(\ref{eq:siglcc}) for $\Lambda \xi^2 \gg1$ the momentum-transfer cross sections reduce to
\begin{subequations}
\begin{eqnarray}
\frac{\bar{\sigma}_{ss^\prime}^{(1)}}{\lambda_\D^2} & \simeq & \frac{4\pi}{(\Lambda \xi^2)^2} \ln (\Lambda \xi^2 )  , \label{eq:sco1} \\
\frac{\bar{\sigma}_{ss^\prime}^{(2)}}{\lambda_\D^2} & \simeq & \frac{8\pi}{(\Lambda \xi^2)^2} \biggl[ \ln (\Lambda \xi^2 ) - \frac{1}{2} \biggr]  , \label{eq:sco2} \\
\frac{\bar{\sigma}_{ss^\prime}^{(3)}}{\lambda_\D^2} & \simeq & \frac{12\pi}{(\Lambda \xi^2)^2} \biggl[ \ln (\Lambda \xi^2 ) - \frac{2}{3} \biggr]  , \label{eq:sco3}
\end{eqnarray}
\end{subequations}
where the next correction to the order unity terms in square brackets are $\mathcal{O} (\Lambda^{-2} \xi^{-4})$. Retaining these order unity corrections, the generalized Coulomb logarithms can be written 
\begin{subequations}
\begin{eqnarray}
\Xi_{ss^\prime}^{(1,k)} &=& \Gamma (k) [ \ln \Lambda + \Psi (k)] , \label{eq:xi1bc} \\
\Xi_{ss^\prime}^{(2,k)} &=& 2 \Gamma (k) [ \ln \Lambda + \Psi (k) - 1/2] , \\
\Xi_{ss^\prime}^{(3,k)} &=& 3 \Gamma (k) [ \ln \Lambda + \Psi (k) - 2/3 ], \label{eq:xi3bc}
\end{eqnarray}
\end{subequations}
in which $\Psi (k)$ is the Digamma function [$\Psi (x) = \Gamma^\prime (x)/\Gamma(x)$]. 

This expansion is useful for retaining an order unity correction to $\ln \Lambda$. Plasma transport theory is often limited to $\sim 10\%$ accuracy because $\ln \Lambda \sim 10$ and order unity corrections are usually omitted. Retaining these leads to a more accurate theory, and allows access to regimes with $\ln \Lambda \lesssim 10$. These order unity corrections have been obtained by others using a variety of techniques that typically rely on renormalization of the collision operator in order to avoid the traditional logarithmically divergent impact parameter integral.\cite{aono:68,lifs:81,brow:05} The calculation of this section provides a physically intuitive, and mathematically simple, way to access this physics. It can be further extended by applying the screened Coulomb potential directly. 

%The usefulness of this expansion is to retain an order unity correction. \textcolor{red}{Add a note about the order unity correction and the connection with other theories. Also add a note that the main physics at the unity order that is missed in Landau's approach is the large angle scattering.} 

%%%%%%%%%
\subsection{Screened Coulomb potential}

Rather than imposing screening in an ad-hoc manner by truncating the interaction range at the Debye length, the screened Coulomb potential can be applied directly.\cite{libo:59,maso:67,paqu:86,khra:02,baal:12} Although this effective potential is also valid only for weak coupling, it provides a formulation that is self-consistently convergent. It is also instructive to consider because it illustrates the importance of large angle collisions and that cross sections are not the same for collisions between like-charges ($q_sq_{s^\prime}>0$) and unlike-charges ($q_s q_{s^\prime} <0$) as the coupling strength increases. In contrast, the Rutherford scattering cross section (valid at asymptotically weak coupling) does not distinguish attractive or repulsive collisions. Furthermore, it provides a simple example to illustrate useful techniques for numerical evaluation of the theory. 

\begin{figure}
\begin{center}
\includegraphics[width=8cm]{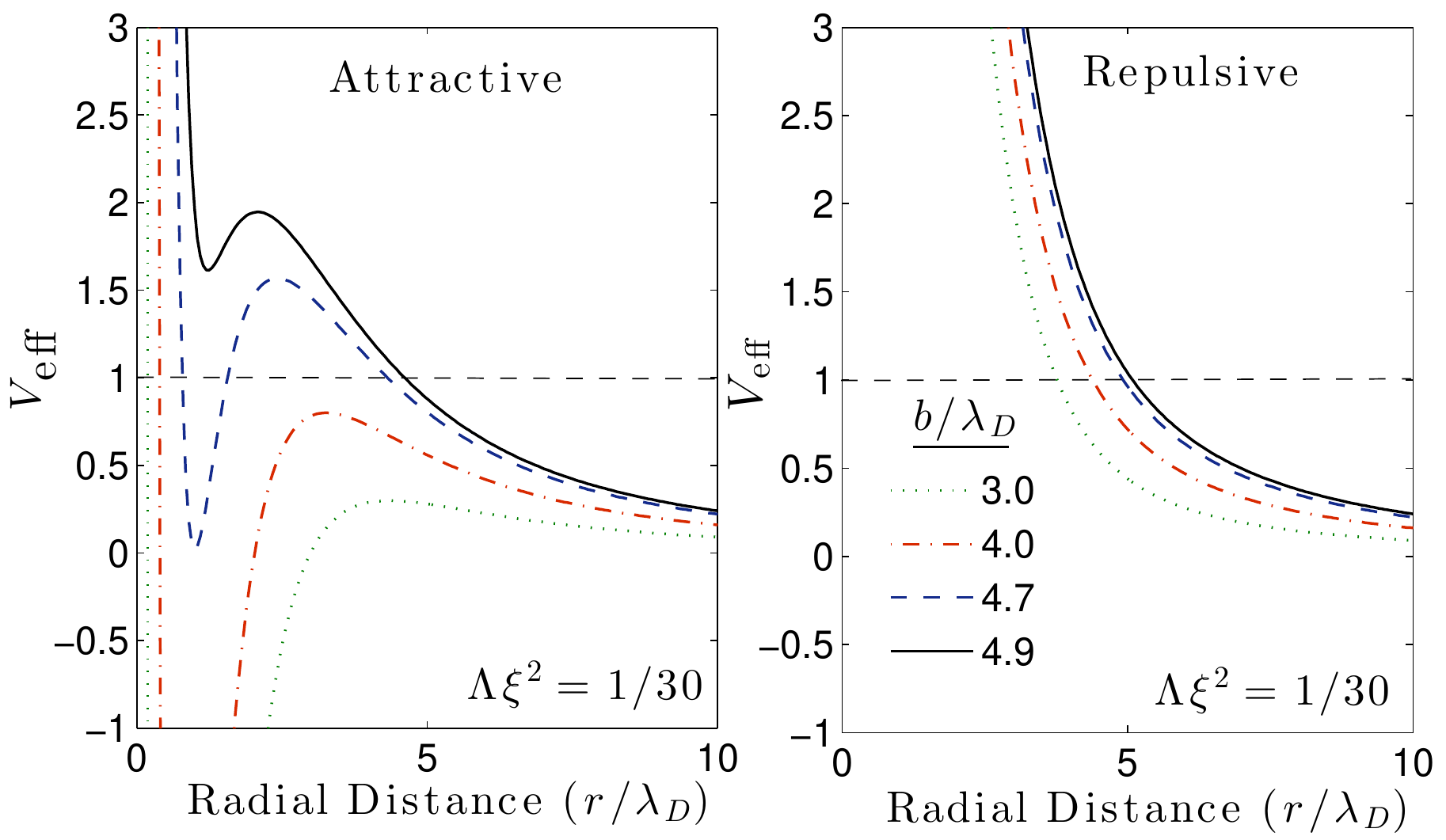}
\caption{The parameter $V_\textrm{eff}$ calculated from the screened Coulomb effective potential for attractive and repulsive collisions. For attractive collisions, three roots (of $V_\textrm{eff}=1)$ are found over a certain range of impact parameters. The distance of closest approach is the largest root.}
\label{fg:veff}
\end{center}
\end{figure}

\begin{figure}
\begin{center}
\includegraphics[width=8cm]{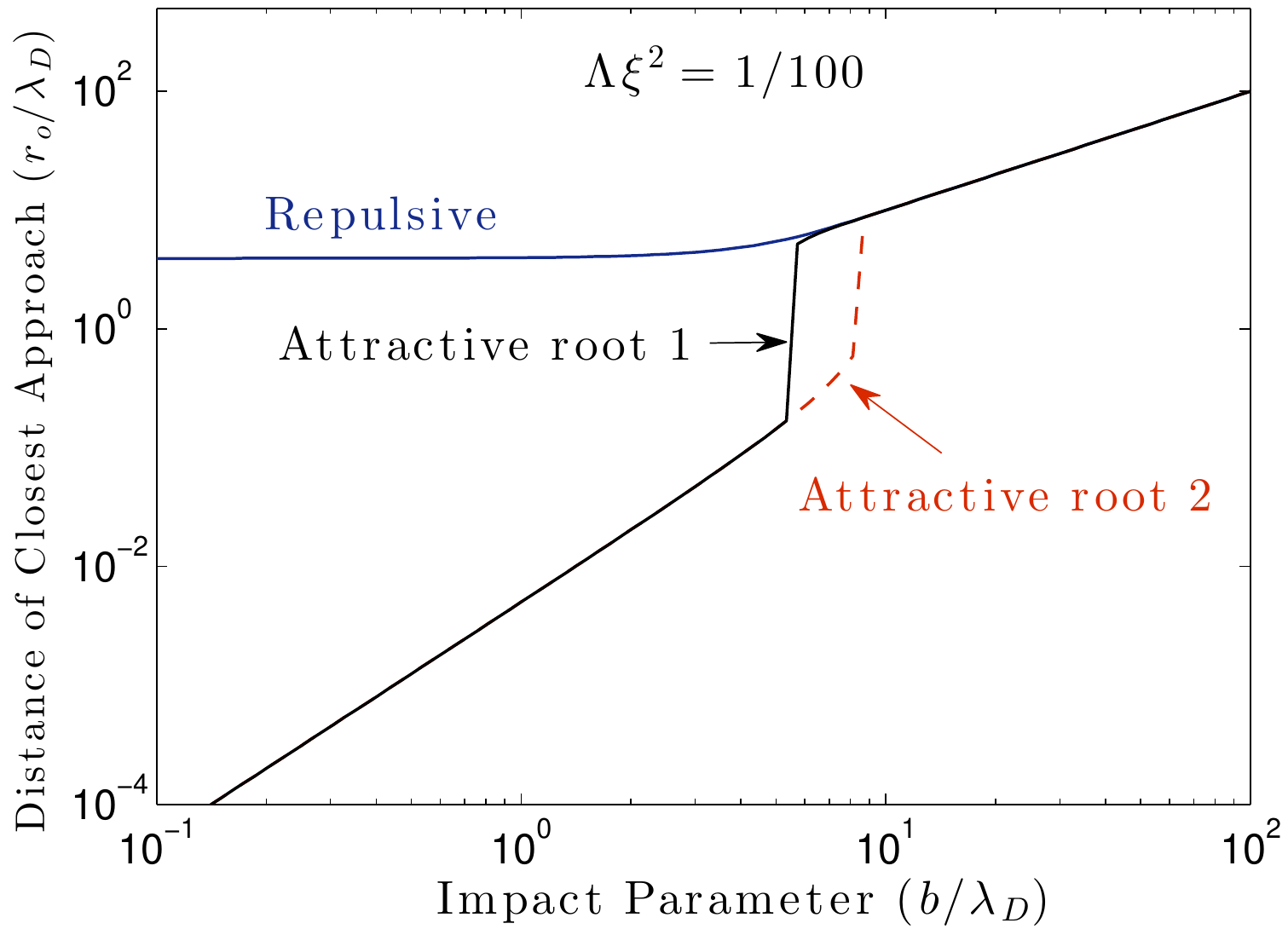}
\caption{Calculation of the roots of $V_{\textrm{eff}}$ from the screened Coulomb potential. Multiple roots can be present for attractive collisions when $\Lambda \xi^2 \lesssim 1$.}
\label{fg:ro}
\end{center}
\end{figure}

Computing $\Omega$ numerically requires evaluation of three imbedded integrals: the scattering angle [Eq.~(\ref{eq:theta})], the momentum-transfer cross section [Eq.~(\ref{eq:sigl})], and the generalized Coulomb logarithm [Eq.~(\ref{eq:xilk})]. The first step in evaluating the scatting angle is to determine the distance of closest approach ($r_o$) from a root of $1 - V_{\textrm{eff}} = 0$ where 
\begin{equation}
V_{\textrm{eff}} = \frac{b^2}{r^2} + \frac{2\phi_{ss^\prime}(r)}{m_{ss^\prime} u^2}  . \label{eq:veff}
\end{equation}
For the screened Coulomb potential $V_{\textrm{eff}} = \bar{b}^2/\bar{r}^2 \pm 2 \exp(-\bar{r})/(\bar{r}\Lambda \xi^2)$ where $\bar{r} = r/\lambda_\D$ and $+$ refers to repulsive collisions (like-charges) while $-$ refers to attractive collisions (unlike charges). Figure~\ref{fg:veff} shows that Eq.~(\ref{eq:veff}) can have multiple roots for attractive collisions. The binary collision picture assumes that scattering particles start asymptotically far apart, so it is important that the numerical routine associates $r_o$ with the largest of these roots. Figure~\ref{fg:ro} shows a range of impact parameters over which multiple roots are found. This property of having multiple roots arises when $\Lambda \xi^2 \lesssim 1$, which corresponds with the onset of strong coupling. Although the next section will show that correlations must also be accounted for in this regime, the properties that attractive and repulsive collisions can be distinguished and that there are multiple roots  of Eq.~(\ref{eq:veff}) persist. In fact, the oscillations found in the effective potential at strong coupling will result in multiple roots even for repulsive collisions. 

The scattering angle integral formally covers an infinite range in the distance $r$. Although convergence is achieved by covering a range determined by the interaction length scale, this length varies broadly as the other integration variables change ($b$ and $\xi$). For this reason, we have found it convenient to use the substitution $w=b/r$ and write Eq.~(\ref{eq:theta}) as 
\begin{equation}
\Theta = \int_0^{w_\textrm{max}} dw [ 1 - w^2 -2\kappa \phi_{ss^\prime} /(q_sq_{s^\prime})]^{-1/2}  \label{eq:thetaw}
\end{equation}
and $w_{\textrm{max}} = b/r_o$. For screened Coulomb $\kappa \phi_{ss^\prime} /q_sq_{s^\prime} = \pm w \exp(-\bar{b}/w)/(\Lambda \bar{b})$. In this case the only independent variable in the momentum-transfer cross section is $\Lambda \xi^2$. Once $w_\textrm{max}$ is determined for a given $b$ and $\xi$, $\Theta$ is determined from Eq.~(\ref{eq:thetaw}). The result is used in Eq.~(\ref{eq:sigl}) and Eq.~(\ref{eq:xilk}) to iteratively solve for the momentum-transfer cross section and generalized Coulomb logarithm. 

\begin{figure}
\begin{center}
\includegraphics[width=8cm]{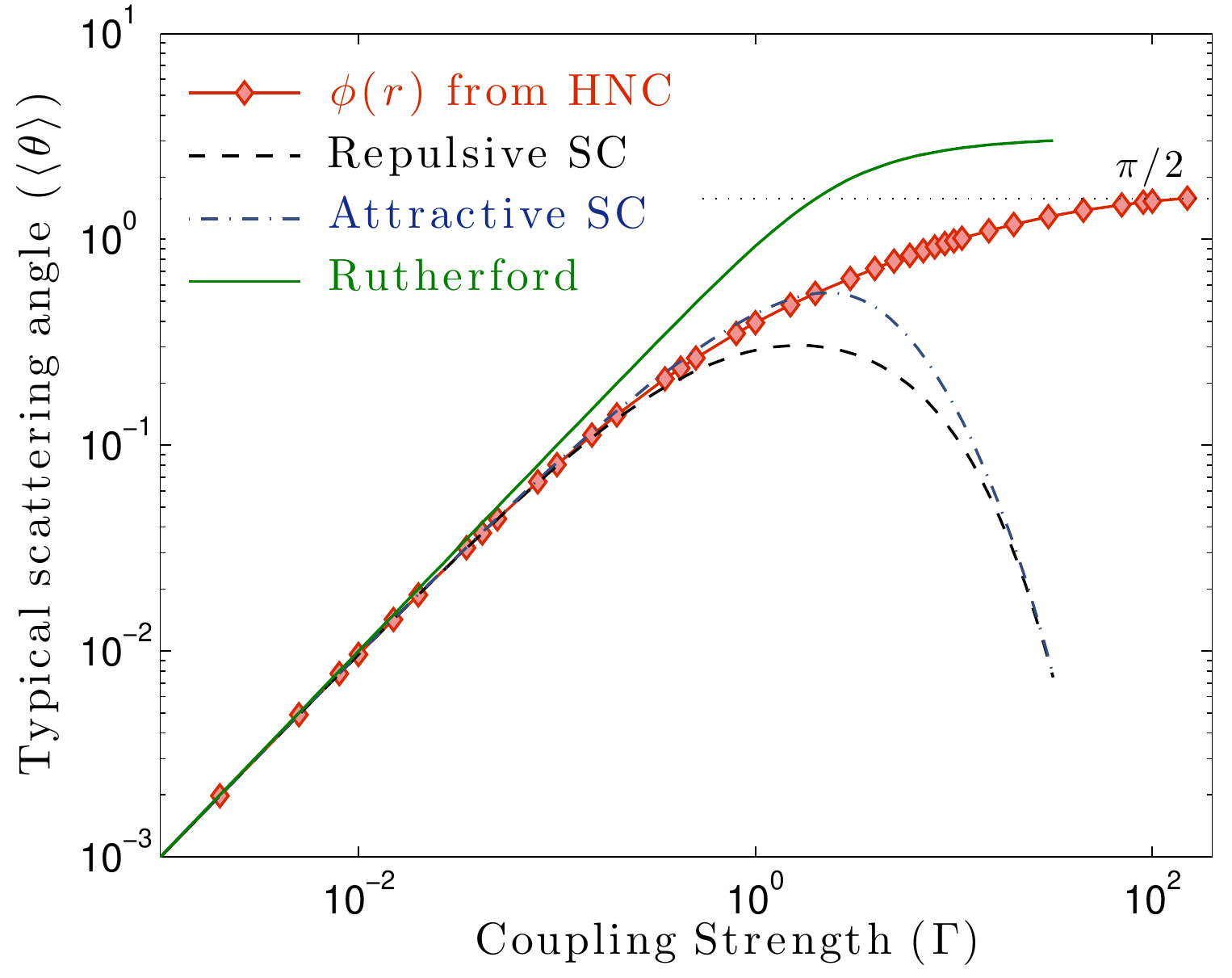}
\caption{The dependence of a ``typical'' scattering angle on coupling strength computed from the screened Coulomb and HNC effective potentials for a OCP. The Rutherford scattering angle is also shown. Here $\la \theta \ra = \theta (b=a, \xi=1)$ and $\theta = \pi - 2 \Theta$.}
\label{fg:theta}
\end{center}
\end{figure}

To better visualize the fact that ``typical'' scattering angles increase with coupling strength, we consider evaluating Eq.~(\ref{eq:thetaw}) for typical parameters $b$ and $\xi$. At any coupling strength, the most probable speed is approximately unity ($\xi \simeq 1$), as the $\xi^5 \exp(-\xi^2)$ factor in Eq.~(\ref{eq:xilk}) demonstrates. On the other hand, the typical impact parameter depends strongly on the coupling strength. To quantify an average impact parameter, we chose the inter-particle spacing $b=a$. Figure~\ref{fg:theta} shows numerical solutions of the scattering angle $\theta = \pi - 2\Theta$ from the different effective potential models choosing $b=a$ and $\xi=1$ [$\la \theta \ra \equiv \theta(b=a, \xi =1)$]. This figure demonstrates the validity of the small scattering angle expansions at weak coupling, but that the approximation breaks down as the coupling strength approaches 1. 

%, which varies from the Debye length in the weakly coupled regime to the inter-particle spacing in the strongly coupled regime. Figures~\ref{fg:gr_r} and \ref{fg:gr_cont} suggest that this distance is the location where $g(r)$ transitions from a small value to 1. This point is discussed further in Ref.~\onlinecite{dali:09}. To obtain a ``typical'' value of $b$ at a given coupling strength, we define $\la b\ra$ as the $r$ location where $g(r) = 0.5$. \textcolor{red}{Show a plot of $\la \Theta \ra$ and $\la b \ra$ and explain the results here}. 

\begin{figure}
\begin{center}
\includegraphics[width=8cm]{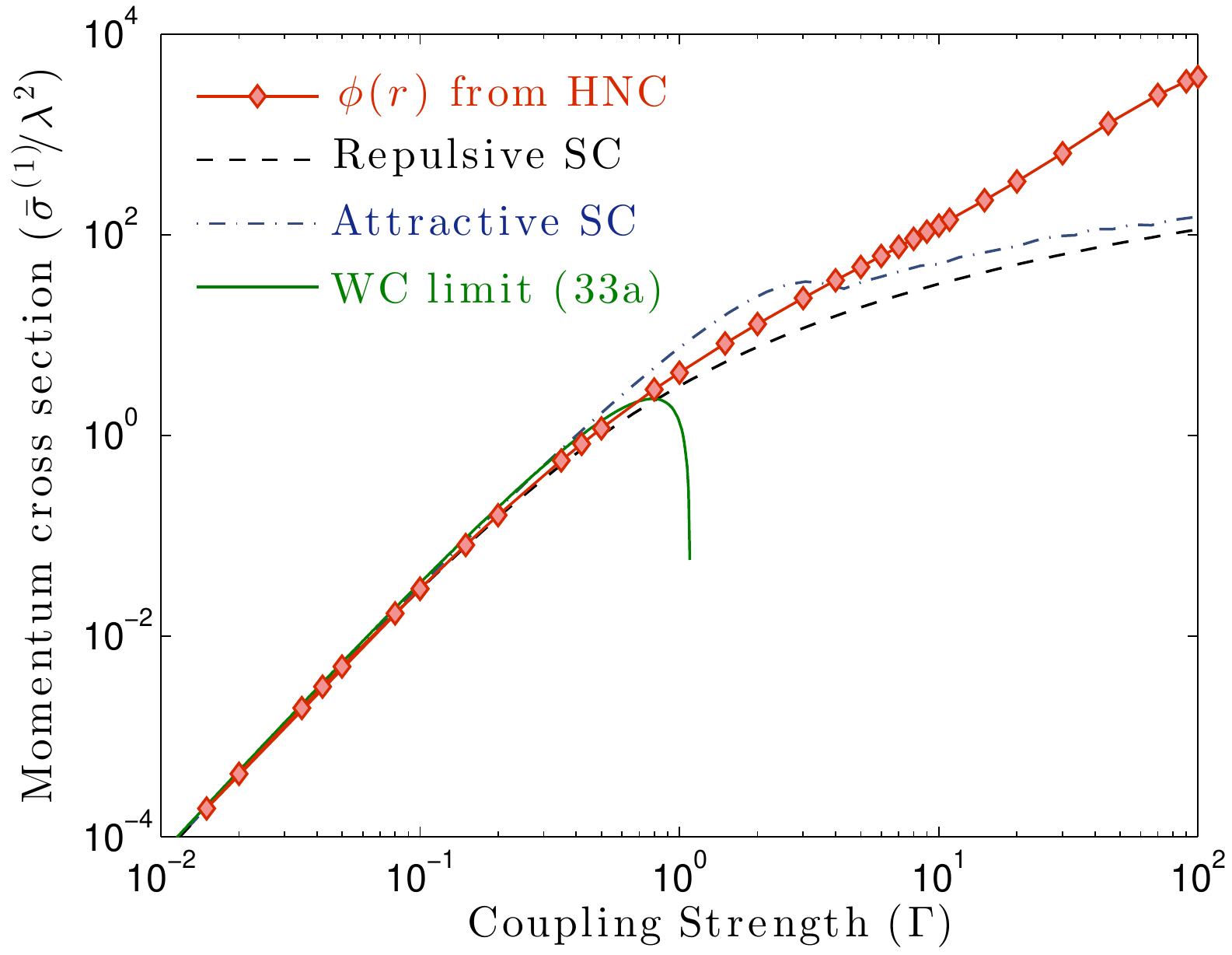}
\caption{First momentum-transfer cross section calculated using screened Coulomb and HNC effective potentials. Also shown is the analytic result obtained for a bare Coulomb potential in the weakly coupled limit [from Eq.~(\ref{eq:sco1})]. }
\label{fg:sig_1}
\end{center}
\end{figure}

\begin{figure*}
%\begin{center}
\includegraphics[width=14cm]{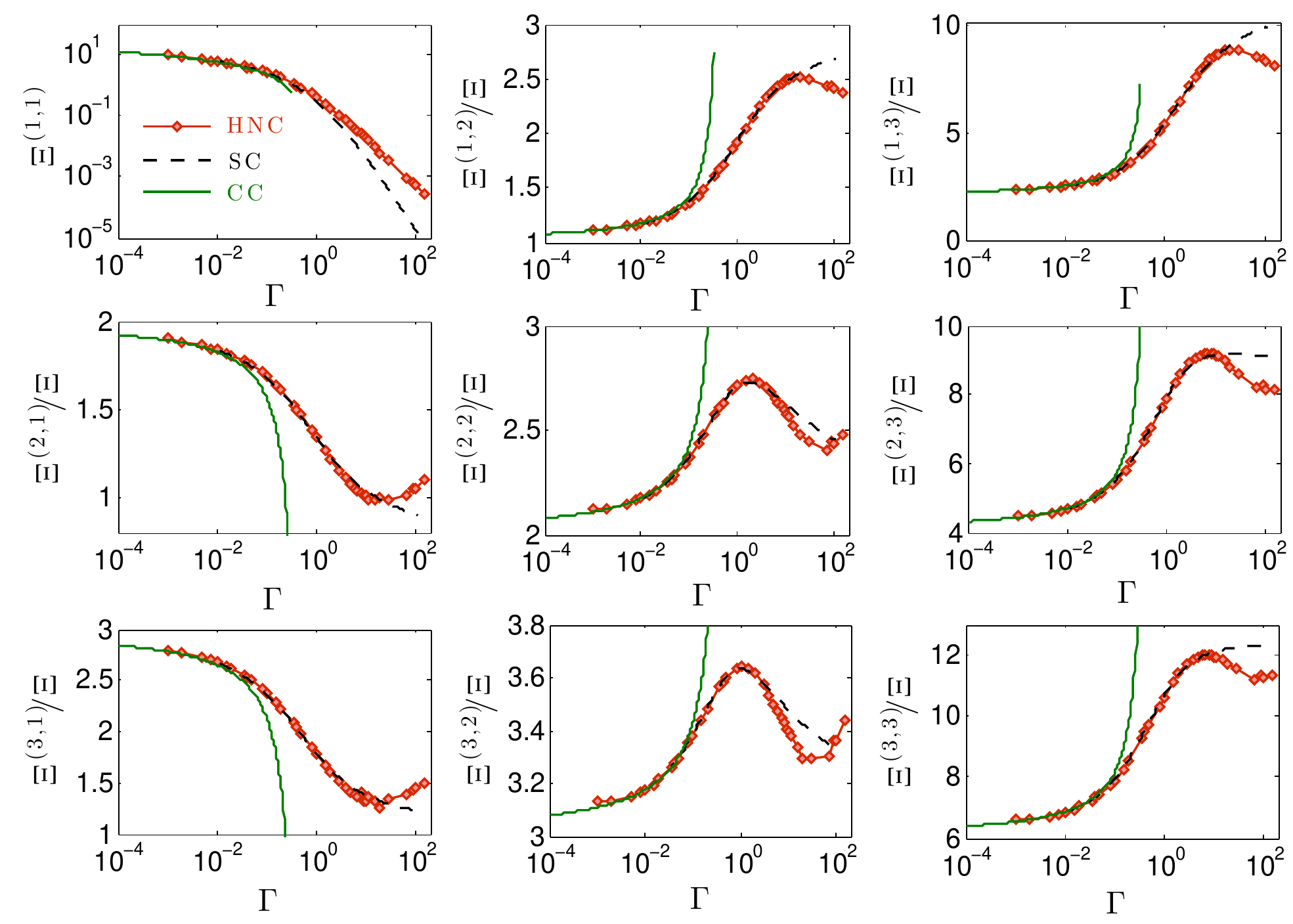}
\caption{Generalized Coulomb logarithms for a OCP calculated using the screened Coulomb (dashed lines) and HNC (diamonds) effective potentials. Also shown are the analytic results from Eqs.~(\ref{eq:xi1bc})--(\ref{eq:xi3bc}) using the cutoff Coulomb potential in the weakly coupled limit (solid lines). Data for $(l,k) \neq (1,1)$ are normalized by $\Xi = \Xi^{(1,1)}$.}
\label{fg:xiocp}
%\end{center}
\end{figure*}

Figure~\ref{fg:sig_1} shows the first momentum-transfer cross section as a function of coupling strength for the attractive and repulsive screened Coulomb potentials. For this plot $\xi = 1$ has been chosen, and the normalization ($\lambda$) is an arbitrary length scale. This figure shows that both solutions converge to the bare Coulomb result of Eq.~(\ref{eq:sco1}) in the weakly coupled limit. The attractive and repulsive solutions are distinct for $\Gamma \gtrsim 1$, but this is also the region where correlation effects onset. Figure~\ref{fg:xiocp} shows the resultant generalized Coulomb logarithms for the repulsive potential. One noteworthy point here is that the ratios $\Xi^{(l,k)}/\Xi^{(1,1)}$ asymptote to constants in the weakly coupled limit, whereas they have a functional dependence on $\Gamma$ at moderate and strong coupling. For instance, these provide the input for the famous Spitzer conductivity problem,\cite{spit:53} where these numbers provide the higher-order corrections to the near-equilibrium expansion. This shows that even for moderate correlation these numbers instead must be functions of the coupling parameter. In the next section, we include correlations in these calculations using HNC.

%\begin{widetext}
%\end{widetext}

%%%%%%%%%
\subsection{HNC potential}

The basics of the numerical integration of Eqs.~(\ref{eq:theta}), (\ref{eq:sigl}) and (\ref{eq:xilk}) are the same for the HNC potential as was discussed in the previous section, only the input file changes. However, it can be significantly more computationally expensive at strong coupling due to the oscillations that arise in the effective potential. This is because one must be sure to choose the largest of the possibly several roots of the denominator in Eq.~(\ref{eq:theta}) in order to determine the distance of closest approach. For the screened Coulomb case, at most three roots arise and they are broadly spaced so it is easy to bracket a single root. However, at strong coupling, the oscillating potential can lead to several closely spaced roots. We have solved for $r_o$ using a bracketed root find, but have found that one must be careful to make the bracket search window small enough to only enclose one root. Since this is the first step in each iteration, it can dominate the computation time when the oscillations are closely spaced. Furthermore, we consider only repulsive collisions in these HNC computations. Calculating attractive effective potentials from the HNC equations is computationally more challenging because $g(r)$ diverges at the origin. Obstacles associated with this can be overcome, and the effective potential theory can be applied equally well to these potentials, but we save this topic for a more detailed discussion in a future publication.

Figures~\ref{fg:theta}, \ref{fg:sig_1} and \ref{fg:xiocp} show that correlations have a significant effect on the scattering angle, momentum-transfer cross section and generalized Coulomb logarithms in the strongly coupled regime. The typical scattering angle approaches $90^\circ$ at strong coupling, demonstrating the importance of accounting for large-angle collisions in this regime. These lead to a much larger momentum-transfer cross section at strong coupling.  Since the array of $\Xi^{(l,k)}$ determine the $\Omega_{ss^\prime}^{(l,k)}$-integrals through Eq.~(\ref{eq:cints}), these provide the only input required to evaluate the transport coefficients in the Chapman-Enskog or Grad fluid equations for the OCP. In addition, the generalized Coulomb logarithms would be expected to remain unchanged from the OCP results for certain classes of multicomponent systems. For instance, the HNC equations, and hence the resultant $\Xi^{(l,k)}$, remain unchanged as long as the charge of each ion species in the plasma is the same. Only a rescaling of the coupling strength ($\Gamma$) would be required. In these cases, the multicomponent effects are all described by the terms multiplying $\Xi^{(l,k)}$ in Eq.~(\ref{eq:cints}).  In the next section, we compare the predictions using the data in Fig.~\ref{fg:xiocp} to compute transport coefficients with molecular dynamics simulations for self diffusion and temperature relaxation rates.

%%%%%%%%%
\section{Comparison with MD simulations \label{sec:compare}}

%%%%%%%%%
\subsection{MD approach}

Molecular dynamics (MD) is a first-principles particle simulation technique that solves Newton's force equation for each particle in the plasma.\cite{fren:02}
In our MD simulations, the particle trajectories are determined by accurately solving Newton's equations of motion with the velocity Verlet integrator in periodic boundary conditions. The latter requires an evaluation of Ewald summations over all periodic cells. To this end, we use the particle-particle-particle-mesh (P3M) method,\cite{hock:88,fren:02} which combines high-resolution of close encounters (which are dealt with using nearest neighbor techniques) and rapid, long-range force calculations (which are computed on a mesh with three-dimensional fast Fourier transforms).

We compute the self-diffusion coefficient $D$ from the Green-Kubo relation
\begin{eqnarray}
D=\frac{1}{3N}\sum_{i=1}^N{\int_0^\infty{dt \langle {\bf v}_i(t)\cdot{\bf v}_i(0)\rangle_{eq}}} \label{DGreenKubo}
\end{eqnarray}
in terms of average over the number $N$ of particles of the time integral of the equilibrium velocity autocorrelation function for each particle velocity ${\bf v}_i(t)$.
In Eq.\ (\ref{DGreenKubo}), the brackets $\big\langle \dots \big\rangle_{eq}$ denote an equilibrium (thermal) average at temperature $T$.
In practice, the velocity autocorrelation function is calculated from the time-discretized expression
\begin{eqnarray*}
\lefteqn{\langle {\bf v}_i(n\delta t)\cdot{\bf v}_i(0)\rangle}\nonumber\\
&&=\frac{1}{N_{\rm sim}+1-n}\sum_{m=0}^{N_{\rm sim}-n}{{\bf v}_i((m+n)\delta t)\cdot{\bf v}_i(m\delta t)}\,,
\end{eqnarray*}
where $n\leq N_{\rm sim}$, $\delta t$ is the time step and $N_{\rm sim}$ the number of time steps. The pair distribution functions shown in Fig.~\ref{fg:gr_r} were computed in a similar manner from the definition in Eqs.~(\ref{eq:g2int}) and (\ref{eq:rhogr}).
Additional details on our MD simulations can be found in Ref.~\onlinecite{MithenPhD}.

%%%%%%%%%
\subsection{Self diffusion of the OCP}

\begin{figure}
\begin{center}
\includegraphics[width=8cm]{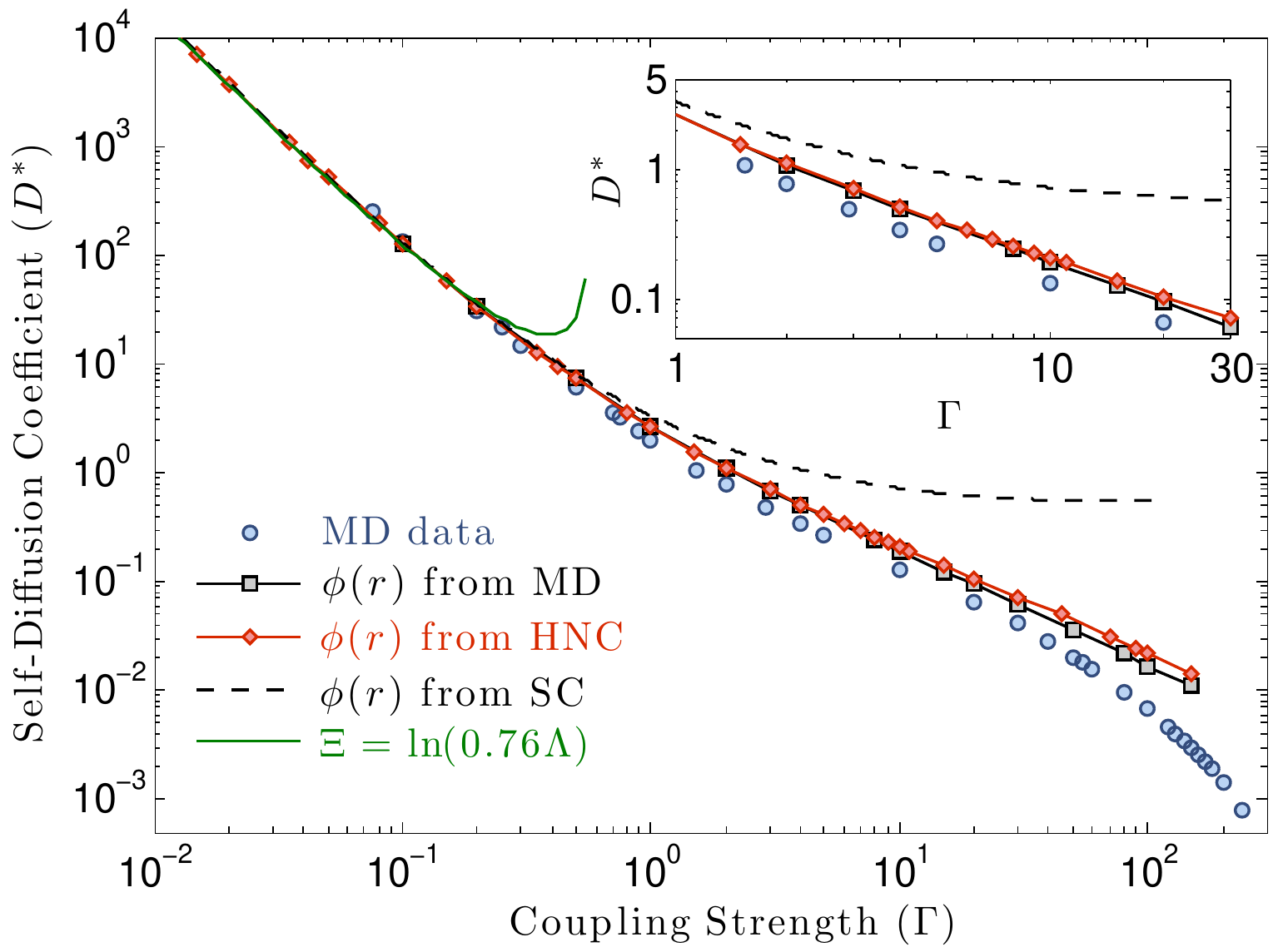}
\caption{Self diffusion coefficient $D^*=D/a^2\omega_p$ of the OCP calculated using classical MD with Green-Kubo relations (blue circles) and using various effective potentials in the Chapman-Enkog collision integrals applied to Eq.~(\ref{eq:d2}): from MD derived $g(r)$ (black squares), HNC (red diamonds) and screened Coulomb (dashed line). This figure is reprinted from Ref.~\onlinecite{baal:13} with permission, Copyright 2013, American Physical Society. }
\label{fg:dstar}
\end{center}
\end{figure}

\begin{figure}
\begin{center}
\includegraphics[width=8cm]{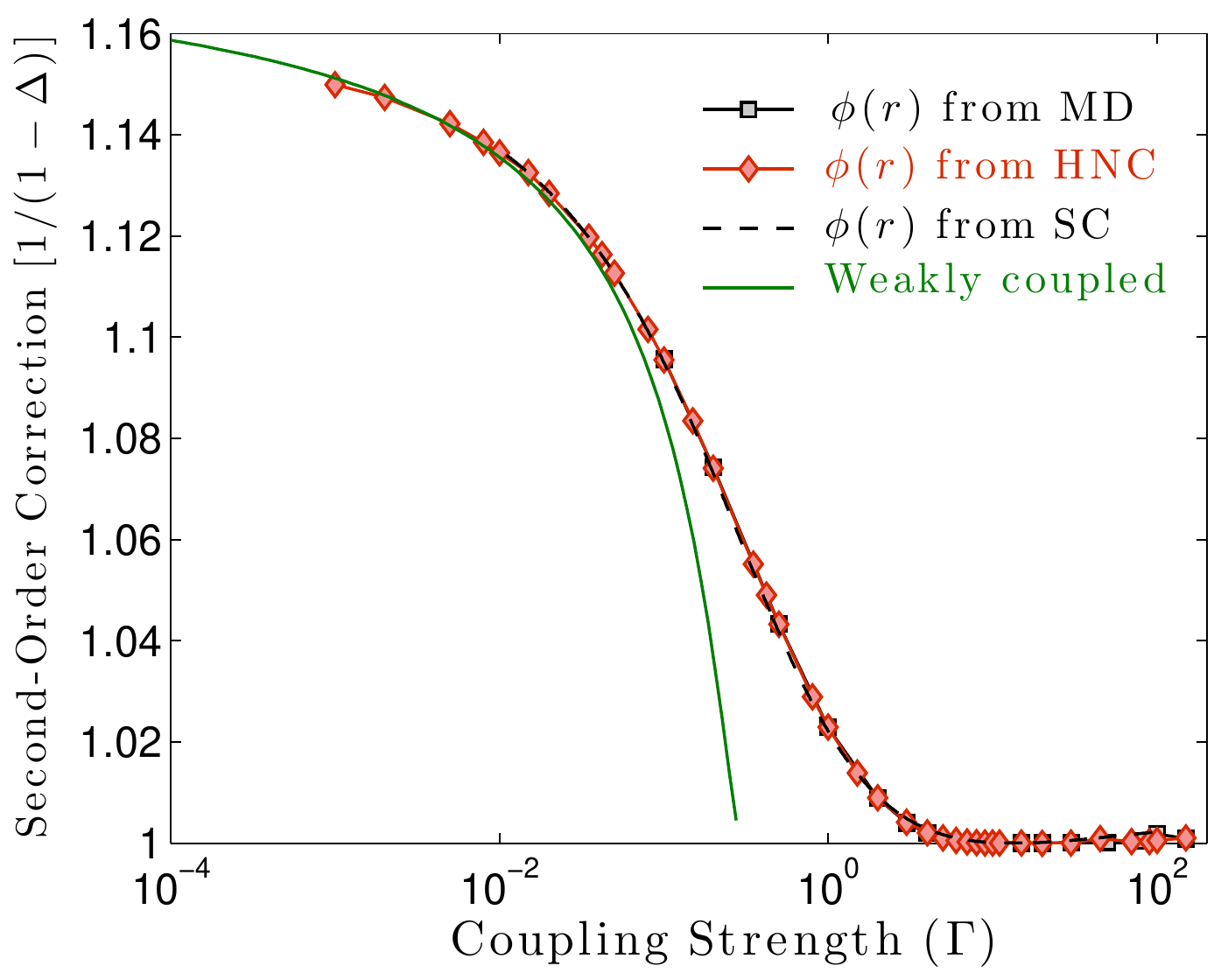}
\caption{Second-order correction to the Chapman-Enskog self-diffusion coefficient of a OCP. The weakly coupled limit asymptotes to 1.2 as $\Gamma \rightarrow 0$.}
\label{fg:delta}
\end{center}
\end{figure}

First, we consider self diffusion in a OCP. The lowest-order mutual diffusion coefficient of the Chapman-Enskog fluid description is\cite{ferz:72} 
\begin{equation}
[D_{ss^\prime}]_1 = \frac{3}{16} \frac{k_\B T}{n m_{ss^\prime} \Omega_{ss^\prime}^{(1,1)}}  .\label{eq:dstar}
\end{equation}
For the OCP ($s=s^\prime$), this can be written as
\begin{equation}
[D^*]_1 = \frac{\sqrt{\pi/3}}{\Gamma^{5/2}} \frac{1}{\Xi^{(1,1)}}  ,
\end{equation}
in which $D^* \equiv D/(a^2 \omega_p)$ and $\omega_p$ is the plasma frequency. Accounting for a second order correction resulting from deviation from the Maxwellian distribution provides\cite{ferz:72}
\begin{equation}
[D_{ss^\prime}]_2 = [D_{ss^\prime}]_1/(1 - \Delta) . \label{eq:d2}
\end{equation}
where 
\begin{equation}
\Delta = \frac{(2 \Xi^{(1,2)} - 5 \Xi^{(1,1)})^2/\Xi^{(1,1)}}{55 \Xi^{(1,1)} - 20 \Xi^{(1,2)} + 4 \Xi^{(1,3)} + 8 \Xi^{(2,2)}} .
\end{equation}

Figure~\ref{fg:dstar} shows a comparison of the calculated self diffusion coefficient obtained by applying the generalized Coulomb logarithms from Fig.~\ref{fg:xiocp} based on the various effective potential models. The figure shows that the effective potential theory indeed provides an extension of the binary collision approximation into the strongly coupled regime. As expected, correlation effects are significant at strong coupling. Thus, the screened Coulomb potential is inadequate in this regime, even though it is convergent. The result computed from the HNC effective potential is approximately 30\% larger than the MD result in the range $1\lesssim \Gamma \lesssim 30$. The comparison between using the HNC calculated $g(r)$ versus the MD calculated $g(r)$ in the theory shows that the slight inadequacies of HNC translates to a negligible difference in the diffusion coefficient. Thus, the disagreement between the model predictions and the MD calculation based on Green-Kubo formula must arise from a break-down of the binary collision picture at sufficiently strong coupling. This regime of coupling strength is known to be a crossover point to liquid behavior in the OCP.\cite{dali:06} This is discussed further in Sec.~\ref{sec:lim}. Figure~\ref{fg:delta} shows that the second order correction to the self diffusion coefficient provided by Eq.~(\ref{eq:d2}) ranges from about $20\%$ in the weakly coupled limit to a negligible level at strong coupling. 
% few percent to a few tens of percent, and is a smaller correction at strong coupling.

%%%%%%%%%
\subsection{Electron-ion temperature relaxation}

\begin{figure}
\begin{center}
\includegraphics[width=8cm]{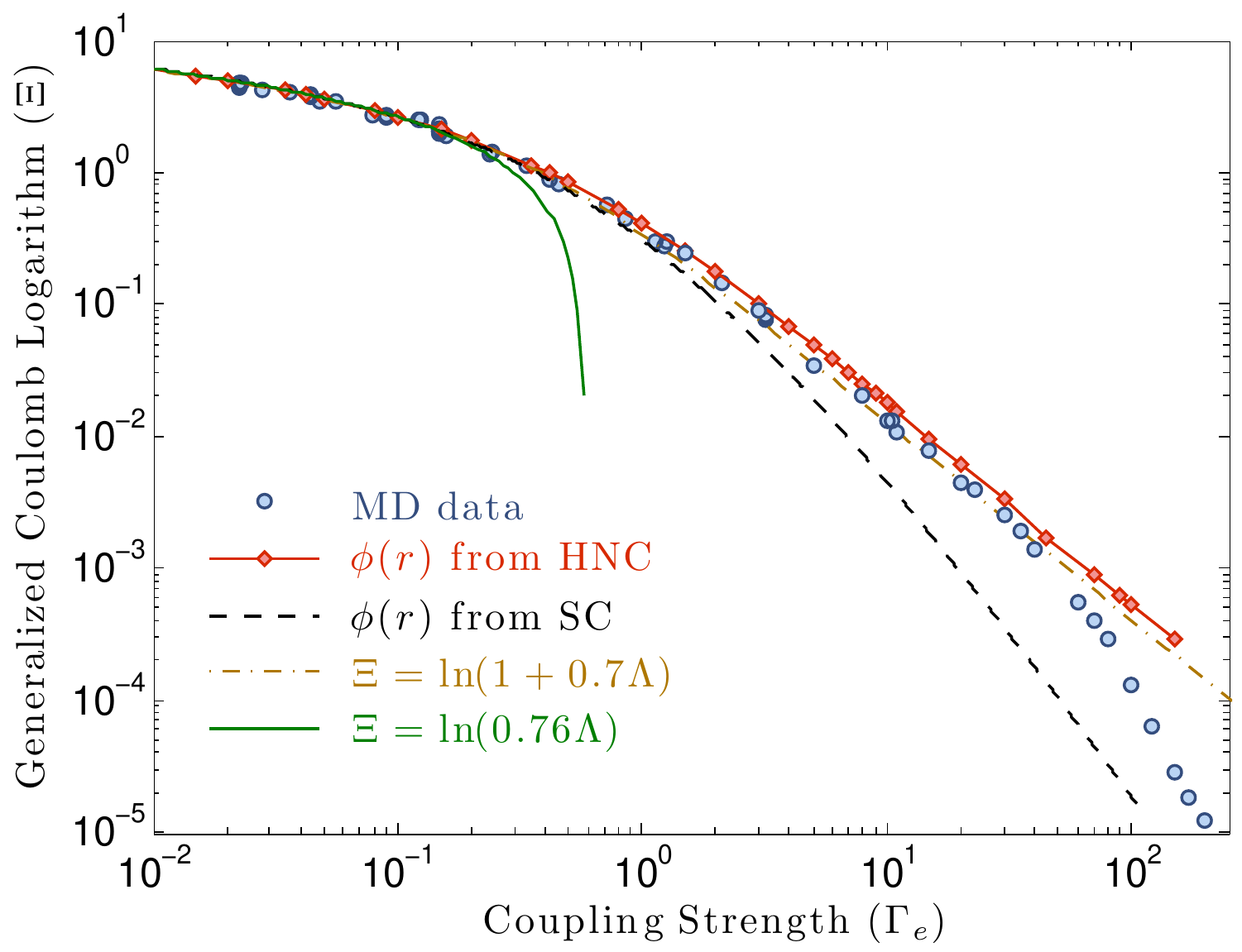}
\caption{The generalized Coulomb logarithm inferred from MD simulations of like-charge electron-ion temperature relaxation compared with predictions of the effective potential theory. This figure is reprinted from Ref.~\onlinecite{baal:13} with permission, Copyright 2013, American Physical Society.}
\label{fg:temp}
\end{center}
\end{figure}

Figure~\ref{fg:temp} shows a comparison between the effective potential theory calculations and MD simulation of temperature relaxation in a like-charged electron-ion plasma. Here like-charge electrons and ions have been used in order to enable an \emph{ab initio} MD simulation. Because classical unlike point charges can come arbitrarily close to one another, these interactions are not able to be self-consistently simulated with MD. Such calculations would require pseudopotentials to resolve these unphysically close interactions for unlike charges, but this is avoided entirely by using like charges. Either case provides a similarly stringent test of the theory. The simulations were performed by analyzing the relaxation of two Maxwellian distributions initiated with different temperatures. The details of these simulations and the analysis technique are provided in Ref.~\onlinecite{dimo:08}. The figure shows the generalized Coulomb logarithm, which for the simulation data was inferred from the temperature relaxation rate from $dT_e/dt = 2Q^{e-i}/2n_e$ where $\mathcal{Q}^{e-i} = -3m_{ei}n_e \nu_{ei} (T_e-T_i)/m_i$ is the energy exchange density from Eq.~(\ref{eq:q2}). The figure shows a similar range of validity of the effective potential theory as was found for self diffusion of the OCP. In addition to these comparisons with MD, the theory was also shown in Ref.~\onlinecite{baal:13} to agree with the most recent experimental data for the velocity relaxation rate of Sr$^+$ ions in an ultracold neutral plasma.\cite{bann:12}

%%%%%%%%%
\section{Limitations and possible extensions \label{sec:lim}}

A fundamental assumption when applying the binary collision approach is that particles start the collision process asymptotically far apart. However, the reality is that potential wells surround particles at strong coupling. These potential wells can trap particles, which creates a correlation in the initial condition between the two scattering particles that is neglected in the basic binary collision picture. As the coupling strength increases, a larger fraction of particles spend most of the time trapped in such wells. Thus, the basic picture of binary collisions breaks down at sufficiently strong coupling. In this regime, the characteristic particle trajectories are better described as jumps from well to well. Transport in this regime is sometimes called caging\cite{dali:06,gold:00,donk:02} and its dominance likely provides the limitation of the binary collision approach at sufficiently strong coupling. 

Another assumption of this theory is that the particles interact though a central and conservative force, implying that that force depends on only the distance between the particles. In reality, particles moving faster than the thermal speed will have a distorted dielectric cloud arising from the wake effect of the particle streaming through a medium. This effect can be significant for the stopping power of fast particles.\cite{grab:13} For the near equilibrium properties discussed in this work, this is expected to be a minor modification because transport is dominated by the vast majority of particles which have speeds slower than the thermal speed (and thus negligible wake effects). However, one could envision extending this approach either by considering dynamic pair distribution functions, $g(r,\omega)$, or modifying the effective screening length along the lines of the approach taken in Ref.~\onlinecite{grab:13}. 

Finally, the transport properties described by the Chapman-Enskog or Grad type theories are based on the kinetics of the particle trajectories. At strong coupling contributions to the transport coefficients arise solely from the potential energy of the particles. For instance, the Green-Kubo relations for viscosity and thermal conductivity have kinetic-kinetic and potential-potential terms (as well as mixed terms). We have found that the effective potential approach fits very well the kinetic-kinetic contributions only. For viscosity this leads to only a slightly narrowed range of validity of the overall theory, but the main distinction is that the theory breaks down in a more dramatic fashion outside of this range. This topic will be discussed in detail in a later publication. 

%%%%%%%%%%
\section{Summary} 

We have found that traditional transport theories based on a binary collision approximation can be extended into the strong coupling regime by applying an interaction potential that accounts for many-body correlation effects. By associating this potential with the potential of mean force, a simple connection can be made to the pair distribution function $g(r)$. This opens the door for a host of approximations for $g(r)$ to provide the input for the transport theory. We concentrated on the HNC approximation, which is a standard approximate method for simple systems. The results showed extension of common transport coefficients, in particular self diffusion and electron-ion temperature relaxation, by a couple of orders of magnitude in coupling strength. This provides a physically intuitive approach that has several practical benefits. These include the computational simplicity and efficiency of the formulation, as well as its versatility in that it fits naturally within the most common approaches to fluid transport modeling, namely the Chapman-Enkog or Grad methods. This theory may provide a computationally efficient means of modeling transport properties in several applications where plasmas cross coupling boundaries without reaching the very strong coupling regimes where the basic assumptions underlying the binary collision approach break down.

%%%%%%%%%%%%%%%%
\begin{acknowledgments}

The work of S.B. was supported in part by the University of Iowa and in part under the auspices of the National Nuclear Security Administration of the U.S. Department of Energy (DOE) at Los Alamos National Laboratory under Contract No. DE-AC52-06NA25396.
The work of J.D. was supported by the DOE Office of Fusion Sciences.

%This work was carried out under the auspices of the National Nuclear Security Administration of the U.S. Department of Energy at Los Alamos National Laboratory under Contract No. DE-AC52-06NA25396.

\end{acknowledgments}

%\bibliography{refs.bib}

\end{document}